# AN ECONOPHYSICS MODEL FOR THE CURRENCY EXCHANGE WITH COMMISSION

Ion SPÂNULESCU[*], Victor A. STOICA[*] and Ion POPESCU[**]

***Abstract:*** *In this paper an econophysics model for the currency exchange operations with commission is proposed. With this purpose some analogies and similarities of the processes that take place in the frame of the electrochemical system made from electrodes sunk into a solution of electrolytes and the process of the currency exchange and determination of the international currency purchasing power have been used. Some contact phenomena at the electrode/electrolyte separation surface, the physical principles of an electrochemical source operation and the determination of the sale attractiveness or the "potential" of the currency that is to be exchanged are also introduced and analyzed.*

## 1. Introduction

In last decade the methods of econophysics – a science recently appeared between the 20th and 21$^{st}$ centuries – have been largely applied in order to model different economic processes, especially from the financial area [1-10], the investments [11-14], or the social phenomena [11-15].

In this work an econophysic model for the illustration of currency exhange and the settlement of the international currencies purchasing power, on the basis of the analogy between the physical processes that takes place in the electrochemical systems and currency exchange operations with commission is proposed.

In order to substantiate and for the illustration of that analogy, in the section 2 there are made some physical considerations upon the electrochemical processes that take place at the contact between different metals, assimilated with the international currencies: Dollars, Euro, Japanese Yen, Pounds etc., and watery solutions with electrolytes (assimilated

---

[*] Bucharest Hyperion University, 169 Calea Călăraşilor, St., Bucharest, 030615
[**] Bucharest Spiru Haret University, 13 Ion Ghica, St., Bucharest, Sector 3

with economic environment in which different exchange currencies with commissions) are practiced.

In the frame of this new econophysics model, the currency exchange is assimilated with the ionic exchangeable process between a watery solution with metallic electrodes which is in contact with, so that the potential of the electrode from the electrode – electrolyte contact represents the sale attractiveness or the economic power of the purchasing currency, from the economic point of view.

In sections 3 and 4, the econophysics model for the currency exchange and for the international currency purchasing power, and the influence of the physical and economic factors upon the currency exchange parameters are introduced and analyzed.

Finally in section 5, the main conclusions are summarized.

## 2. Contact phenomena. Physical processes in electrochemical systems made of electrodes and electrolytes solutions

### *2.1. Contact phenomena*

Contact phenomena appear both between two solid bodies, which are two different conductors (or semiconductors), and to the contact between a metal (or semiconductor) and a watery solution in which certain substances (salts or acids), named electrolytes, have been put in. To the contact between two different substances appears a contact difference potential $V_c$ as the result of the difference between the work function of the electrons from metals (or semiconductors) of a different type brought into an intimate contact.

As it will be shown afterwards, such a potential "leap" appears also to the contact between a metal (or semiconductor) characterized by the work function $W_m$ and a solution with electrolytes with the work function $W_l$ (Fig. 1,b). In this case the leap (the potential difference) is named as electrode potential. In the present work the electrode potential from an electrochemical system – assimilated with a circuit of a currency exchange – is assimilated with the "potential" of the exchangeable currency, or with the **sale attractiveness** of that currency.

Electrolytes solutions (that contain electrolytes) are part of the category of so-called conductors of $2^{nd}$ type at which the running electrical current is made with the help of the material ions in solutions, not through the electrons, like in the solid conductor bodies.

In the case of electrochemical piles (Volta cell, the Daniell's cell etc.) or the phenomena of electrolysis, galvanization etc., there are usually used pairs of solid conductor electrolytes, partially sunk in acid solutions or alkaline electrodes, and between the two electrodes there appears a potential difference given by the difference between the electrode potential of the two electrodes [16].

## 2.2. The metal-electrolyte contact. The electrode potential (the "potential" of currency or its sale attractiveness)

In our considerations, a foreign currency, that is under interaction with the financial economic environment, is assimilated with a metallic ar semiconductor electrode placed in a system with electrolytes (watery solution) that represents the "economic environment" (Fig. 1,a).

As it has been mentioned, the difference of the contact potentials (currency sale attractiveness) and, respectively, of the potential difference (of the difference regarding the attractiveness of the exchanged currency) does not appear only at the contact between two solid bodies (it does not appear merely at the contact of two currencies), but also at the contact between a solid (a currency) and a liquid (an economic environment), between which certain chemical processes assimilated with processes of transformation or currencies exchange there are in progress (Fig. 1,b). For instance, if a zinc plate (assimilated with a hard currency: the dollar) sink in a watery solution of sulfuric acid ($H_2SO_4$) (that means that the currency is transformed in an investment, $I$, therefore, in an economic variable, (product or service etc.) and under the action of the acid solution zinc starts being corroded and dissolved in the electrolyte (Fig. 1,a), therefore, the currency starts being consumed and transformed in fix funds (see also Fig. 3).

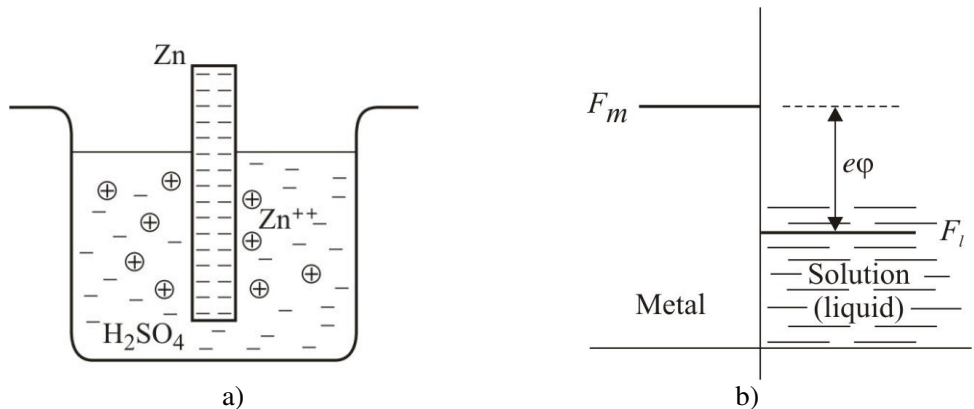

**Figure 1.** Transformation processes in an electrochemical system.

In the case of the figure 1, from physically point of view, under the action of the acid solution, the zinc starts being corroded and dissolved in the solution with electrolyte (Fig. 1,a). In that solution do not pass neutral atoms of Zn, but ions $Zn^{++}$ only; therefore the solution shall be positively charged and the zinc electrode shall remain negatively charged. Thus, between the electrode of Zn and the solution there appears a difference of potential $\varepsilon$ of which value is determined by the value of the "work functions":

$$W = e\varphi \qquad (1)$$

that represents the difference between the chemical potential (the Fermi level, $F_m$) of the ions from the metallic network (of the Zinc) and chemical potential $F_l$ of the ions from the solution (Fig. 1,b):

$$e\varphi = F_m - F_l. \qquad (2)$$

The sign of the electrode potential depends on the nature of the electrolytic solution and electrodes which can have a chemical attractiveness higher or lower than the electrolytic solution, if the Fermi level, $F_l$ (of the solution) is higher or lower than the level $F_m$ (of the metal).

In the case of the zinc, in normal temperature and pressure conditions, for solutions of $H_2SO_4$ normally diluting, this potential is equal with –0,50V. For other metals, the values of this potential differ from metal to metal, being able to take also positive values in accordance with the type of the metal, electronegative or electropositive, in relation to the electrolyte solution. For instance, for a Cu electrode in a sulfuric acid solution the electrode potential is equal with + 0,61V [16].

The electrode potential can be measured only in comparison with the potential of another (metallic electrode) one that is necessarily to be sunk into solution in order to have a closed electrical circuit. This is achieved using an electrochemical chain known as an electrical pile (power source), formed from the electrode of which potential must me measured and a comparison electrode. As comparison electrode has been selected, the normal hydrogen electrode for which, conventionally, a zero value for its normal potential is taken. Thus, the electrode potentials determined in report with the normal electrode of hydrogen represent **relative values**, and they are displayed under this form in different tables (see Table 1) [16].

**Table 1**

The standard normal potentials for different electrodes in report with
the standard potential of the hydrogen electrode (Source: [16])

| | | | |
|---|---|---|---|
| Li/ Li$^+$ | –3,04 mV | Co/ Co$^{2+}$ | –0,28 mV |
| K/ K$^+$ | –2,92 mV | Ni/ Ni$^{2+}$ | –0,23 mV |
| Ca/ Ca$^{2+}$ | –2,87 mV | Sn/Sn$^{2+}$ | –0,14 mV |
| Na/ Na$^+$ | –2,71 mV | Pb/ Pb$^{2+}$ | –0,13 mV |
| Mg/ Mg$^{2+}$ | –2,37 mV | **H$_2$/ 2H$^+$** | **±0,00 mV** |
| Mn/ Mn$^{2+}$ | –1,18 mV | Cu/Cu$^+$ | +0,34 mV |
| 2H$_2$O/ H$_2$+ 2OH$^-$ | –0,83 mV | 2Hg/ Hg$_2^{2+}$ | +0,79 mV |
| Zn/ Zn$^{2+}$ | –0,76 mV | Ag/ Ag$^+$ | +0,80 mV |
| Cr/ Cr$^{3+}$ | –0,74 mV | Hg/ Hg$^{2+}$ | +0,85 mV |
| Fe/ Fe$^{2+}$ | –0,56 mV | Pt/ Pt$^{2+}$ | +1,20 mV |
| Fe/ Fe$^{3+}$ | –0,44 mV | Cl$_2$/ 2Cl$^-$ | +1,36 mV |
| Cd/ Cd$^{2+}$ | –0,40 mV | Au/ Au$^+$ | +1,50 mV |
| Ti/ Ti$^{2+}$ | –0,34 mV | F$_2$/ 2F$^-$ | +2,87 mV |

In the case of a chemical source largely used as there is the Volta cell or the Daniell's cell (Fig. 2,a) the two metallic electrodes are used, namely an electrode of Zn – which represents the anode, or the positive electrode – and a Cu electrode, which represents the cathode, whereto the positive ions from solution come.

The two electrodes are sunk into a watery solution of H$_2$SO$_4$ (the electrolyte), forming the electrochemical chain (Fig. 2,b):

$$Zn \mid H_2SO_4 \text{ (aqua)} \mid Cu \qquad (3)$$

in which the straight bars from (3) represent the separation surfaces of the two phases: solid and liquid.

The electromotive force whide is generated by the pile in open circuits is given by [16]:

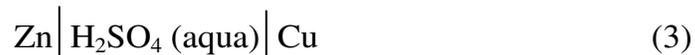
$$e = \varepsilon_+ - \varepsilon_- \qquad (4)$$

in which $\varepsilon_+$ and $\varepsilon_-$ represent the potential leap (contact potentials) from two electrodes, the positive one, of Cu, and the negative one, of Zn (Fig. 2,a) calculated in report with the standard potential (of the hydrogen electrode).

Taking into account the measured values for the electrode potentials of Cu and Zn (Fig. 2,c), the following value is given for the Volta battery voltage [16]:

$$e = -0.61 - (-0.50)V = 1.11\,V \qquad (5)$$

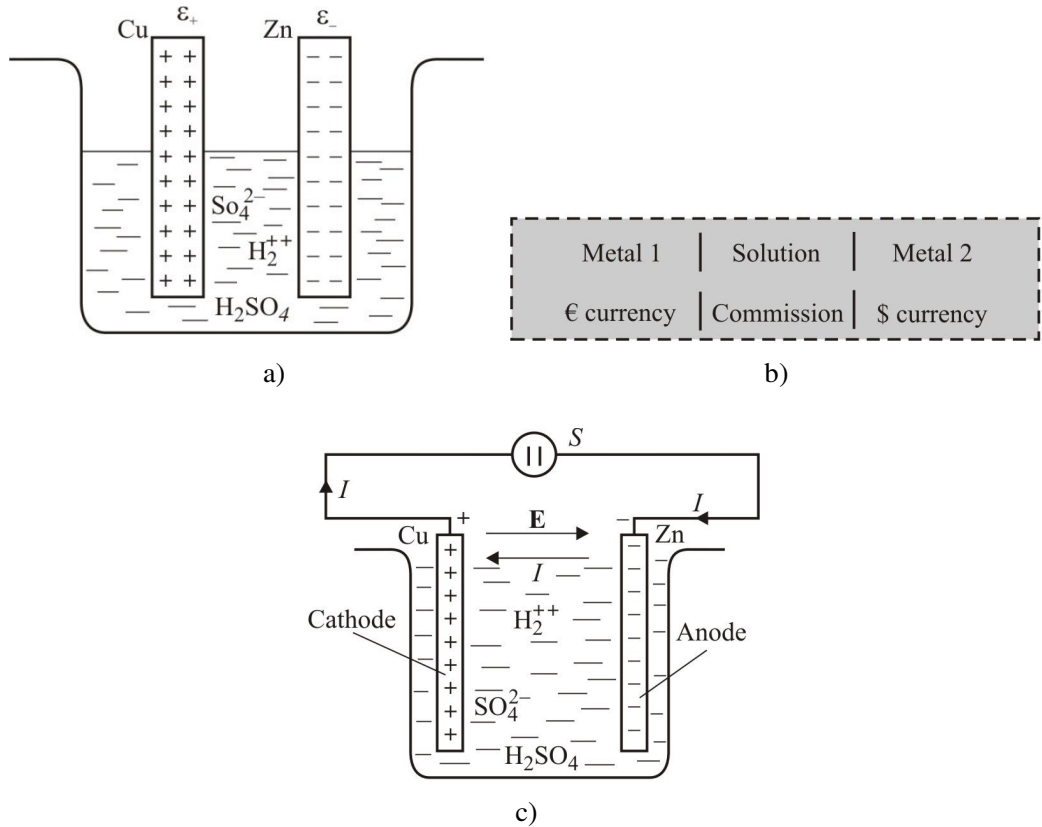

**Figure 2.** The scheme of an electrochemical pile.

## 3. The currency exchange operations modelling using the physical processes from an electrochemical system

The currency exchange is the financial operation through which a currency is exchanged with another one at an exchange office or at banks, for a commission accepted by both partners: the beneficiary, respectively, the clerk of the bank or of the exchange office. The exchange takes place starting from the official parity rate of currency established by the National Central Bank.

Starting from the convention that an electrochemical source or pile is in report of similitude with the currency exchange operation, the metal – electrolyte contact is assimilated with the assembly currency – commission, and potential of electrode with the sale attractiveness of the currency that is determined by the international purchasing power of that certain currency. Under this hypothetical situation, the contact potentials differrence (the sale attractiveness of the currency) and, respectively, the potential difference (the difference between the attractiveness of the currency that is to be exchanged into another currency) do not appear only in case of a contact between two solid bodies (that means between two different currency) but also at a contact between a solid one (international, national currency) and an electrolyte solution one (an economic environment), in the frame of which certain chemical processes, or transformation, are recorded. As it has already been mentioned in the previous section, if a Zn plate (equivalent with a dollars amount) is sunk into a solution of sulfuric acid ($H_2SO_4$) (that is in an economic environment), then that is transformed in an investment, $I$, that is an economic value (quantity), and under the action of the acid solution (through the scroll of the investment process), the zinc (the amount of dollars) starts being corroded and dissolved in electrolyte (Fig. 1,a). Due to the fact that it has been corroded, the Zn electrode is consumed, so that the dollar amount being transformed in fix funds of which value is determined by the height of the potential barrier, $e\varphi = F_m - F_l$, that represents the investment efficiency as a result of currency (dollars) utilization in the economic process (Fig. 3 and Fig. 1,b).

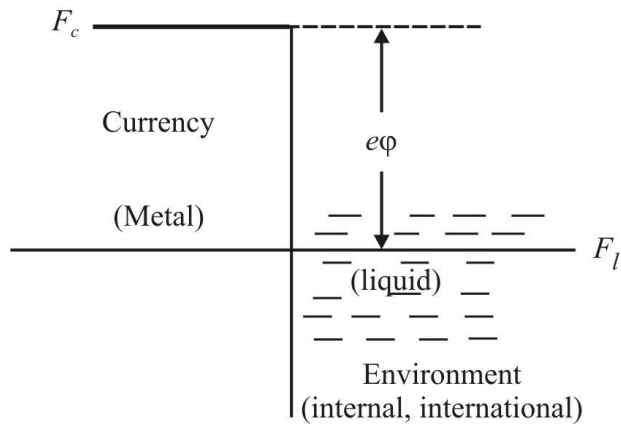

**Figure 3.** The evolution of the currency exchange operation.

As it has already been mentioned, the neutral Zn atoms (certain subdivisions of a currency) do not pass into solution, but only the positive ions $Zn^{++}$ pass (that means discrete subdivisions of dollar, so, cents) and consequently, the solution (the economic environment), shall be positively charged (with fixed funds), and the Zn plate (the amount of dollars)

21

remains negatively charged (it clears out, with the interests to be paid). Thus, between the electrode of Zn (the amount of dollars) and solution (the economic environment) appears a potential difference (a difference between the investment profitableness and the interest rate – that is profit, $p$), respectively an electric field $\vec{E}$ (investor-bank relations) orientated from liquid (investment) toward electrode (the dollar borrowed amount reconstitution and interest one) [16].

The "contact" electric field (investor-bank relations) shall stop further crossing of the metal ions (cents) in the economic environment (the investment being closed). Thus, a dynamic equilibrium is achieved, when for a certain potential difference $\varepsilon$ (the difference between the investment profitableness and the interest rate – that is the profit, $p$), so that, for a certain value of the "field" (the loan to be repaid and the left interest mass), the number of metal ions (cents) that pass into solution (into the economic environment) becomes equal with the ones that turn back to metal (loan repaying plus interests) and the dissolving is stopped (the loan plus the interest are paid). The value of this potential, named electro-chemical potential or electrode potential (assimilated with the international currency purchasing power, that means with its sale attractiveness) depends on the nature and properties of the metal (currency), of the solution (the economic environment which generates it) and of the initial concentration of the metal ions (the external debt of the currency issuing country) in solutions (the currency weight in the national economic Gross Domestic Product – GDP).

In the case of the zinc, in normal temperature and pressure conditions (in conditions of prices normal stability), for solutions of $H_2SO_4$ normally diluting (that means for normally developed economic environments), this potential (international currency purchasing power) is equal with –0,50V for the Zn electrode (see §.2.2 and (5) relation) and (conventionally) equal with 5% on a scale from 0% to 10%, for example, for the international currency purchasing power.

For other metals (currencies), the values of this potential (international currency purchasing power) differ from metal (currency) to metal (currency), being able to take also positive values in accordance with the type of the metal (that certain currency), electronegative (when the currency of a country is not in demand) or electropositive (when the currency of a country is highly in demand), in relation to the electrolyte solution (international economy).

For a Cu electrode (herein considered as Euro) in a sulfuric acid solution (international commerce), it has been shown that the potential of electrode (the equivalent of the international purchasing power) is equal with + 0,61V, from the physical point of view (see §.2.2), but in the case of the currency exchange the international purchasing power is given by the currency parity modification between moments $t_0$ and $t_1$, and between currency values $v_1$ and $v_2$, that means that the currency has a relative value, reported to the comparison currency.

The difference between of the two potential electrodes (the difference between the purchasing powers of two currencies) from the interface of the two phases metal/ solution (currencies / international commerce), appears due to the fact that the chemical potential (the Fermi level) (that means the variation of the international prices) of the ions from the metallic network (of the currencies monetary subdivisions) differs from the chemical potential of the ions in the solution, through the size $e\varphi$ (Fig. 1,b and Fig. 3). The ions (the currency) from environment (intern or international) with a higher chemical potential (with the index of the higher prices) pass into the environment (intern or international) with a lower chemical potential (with the index of the lower prices).

In the case of figure 1,b, for the zinc electrode (dollar currency in the figure 2,b), the metal ions pass into the solution (the cents that pass into the internal market) and the metal (dollar) is negatively charged (it increases its purchasing power) in report with the solution (in report with the international or internal markets).

If pairs of metals-solutions are chosen (currencies – internal markets, pairs) so that the ions from the solution (the subdivisions of the currencies on the internal markets) are at the level of the higher chemical potential (it is on the markets with higher prices index), they shall leave the solution (internal markets) and deposit on the metal (that is the currency mass on the external markets is growing) and that is why the metal shall positively charged itself (the total currency mass grows) in report with the solution (on the international market). The metals (the currencies) with low chemical activity are part of this group (from the internal markets with low growths of the prices indexes) as there the metals with physical equivalent Au, Ag, Pt, Cu, Pa, Hg etc. or with the currencies: euro, pounds, dollar, Japanese yen, Australian dollar, rouble etc. [16]. Thus the sign of the electrode potential (the appreciation/ the depreciation of the currency) is caused of ions tendency whether to leave the metal or on the contrary, to pass from the solution on the metal, is determined by the tendency of the prices to depreciate the currency or to appreciate it [16].

From the ones previously mentioned we may conclude that it is obviously that the potential of the electrode (the international currency purchasing power) is the consequence of a transfer reaction of the ions from the inner part of a phase into another one (that means that the purchasing power is the consequence of an index difference of the internal prices reported on the international market prices ones); On the other hand, as it has been shown, because the galvanic potential ε of a singular electrode cannot be measured, also the international purchasing power of a single currency cannot be measured. As the electrode potential may be determined making an ensemble as an electric pile from the electrode of which potential must be found, and a comparison electrode (Fig. 2,a,b,c); it is the same situation for the international currency purchasing power that can be determined making an ensemble of **two currencies**, one of which international purchasing power must be found, out and one of comparison.

As electrode of comparison (comparison currency) it has been convened that the normal hydrogen electrode (up to the present, the dollar currency), which the 0 value was given to (the value of 5 for dollar, in the convention made by us for currencies), as its normal potential (for its normal international currency purchasing power).

As it has been shown, the normal potentials determined in report with the normal electrode of hydrogen are not absolutely potentials, but **relative values**, $E_0$, that are also displayed in different tables similarly to Table 1 for the electrode potentials of some physical electrodes. In the same way there is concluded that the values of the international purchasing powers of the currencies in report with the dollar, are not international purchasing powers of the singular absolute values, but relative values, being also shown in different financial tables (similar to Table 1).

## 4. The currency exchange modelling through electrolysis process and galvanization with a power generator (an equivalent of a currency source)

The act of currency exchange can be assimilated also with the process of electrolysis, in the case when currency resources supplying, taken from a generator (currency source) similar to the electrolysis process. The electrolysis represents an exchange process (exchange currency) during which the electric power that is supplied by an external source $G$ (Fig. 4,a) is transformed in chemical power that brings contributions to the electrochemical reactions [16].

The reaction takes place in watery solution (economic environment) with electrolytes, (the parity of the two currencies being fixed by the Central Bank). The ions (the flux of the currencies) must circulate free (in accordance with the request-offer report of curency), in the solution with electrolytes (the economic environment). The two electrodes (the currencies) must be tied through an electrolyte with salts (the commission) and a power generator (Central Banks, the issuing authority of the two currencies) assimilated with the stocks of euro and dollars (Fig. 4,b). In this case, the exchange currency operation results may be evaluated through the assimilation of this exchange with the exchange processes that takes place in a system in which the electrolysis takes place, that means in an electrolytic bath that may be with passive electrodes (for example Pt) for the electrolysis of water or of another solution, or with "soluble" anode used in the processes of mettalic covering (galvanization) like nickel coating or chromium plating etc.

In the case of nickel plating, in an electrolytic bath, Ni sulphate as electrolyte is used, the anode is a nikel plate, and to the cathode the metallic part (or piece) that must be nickel plated is attached (Fig. 4,a).

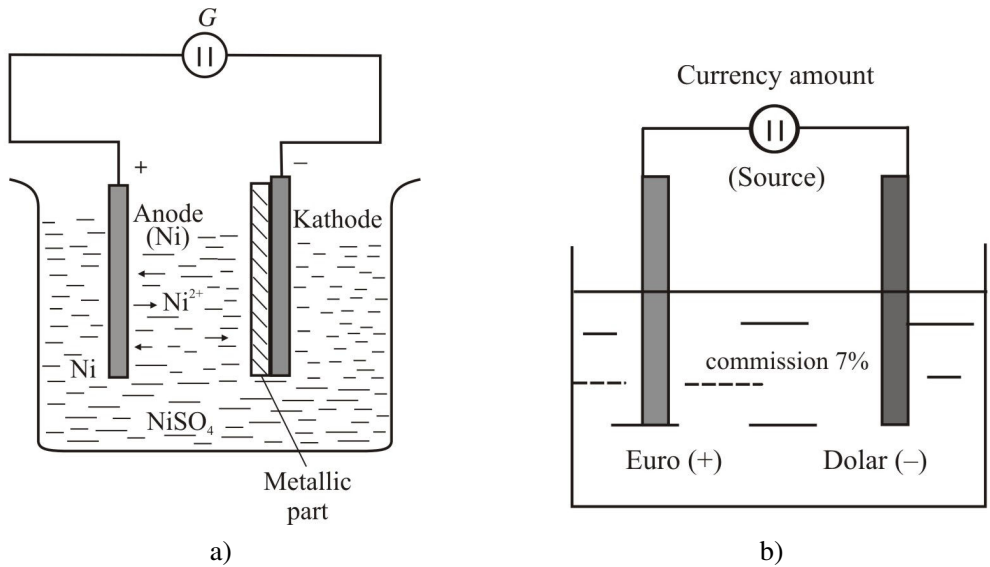

**Figure 4.** Equivalent structures in the case of currency exchange modelling as currency source (currency salvo).

Under the action of the electric power supplied by the power source $G$, the Ni sulphate is dissociated in $Ni^{2+}$ and $SO_4^{2-}$, and the resulted ions

25

head for the electrodes from the bath where the following chemical reactions are in process [16]:

To cathode:
$$Ni^{2++} + 2e = Ni \qquad (6)$$
Nickel is deposited (until the process ends);

To anode:
$$Ni(plate) + SO_4^{2-} = NiSO_4 + 2e^-. \qquad (7)$$

Two electrons ($2e^-$) are given in and the nickel sulphate $NiSO_4$ is recuperated and it passes in the solution of which concentration stays thus constant.

In the case of a currency exchange between two currencies (Euro and Dollar, for example) there may be noticed some considerations similar to the ones of galvanization (chromium and nickel plating etc.). Thus, if an electric power (a stock of currencies) is applied between two electrodes (the currencies of euro and dollars), the positive ions (eurocents) migrate to the cathode (the petitioner of euro), while the negative ions (the American cents) shift to the anode (the petitioner of dollars). The positive ions (eurocents) are named cations (euro) while the negative ions (the American cents) are named anions (dollars). The cations (euro) are able to capture electrons (a growth of the official quota), on the power of their valence (the official quota) similar with the relation (6) for the reaction from cathode where the metal deposit is achieved (Ni, in the case of the figure 4,a). The anions (the dollars) react oppositely (they depreciate); in case of contact (exchange) with the anode (euro) they shall give in their electrons (they suffer a diminution of the official quota) in order to take a steady position and become a stable element (a stable dollar) similar to the relation (7), for the reaction from the anode (Fig. 4). To the cathode (currency – that is the dollar) the cations (euro) are reduced (they diminish the exchanged quantity) and the anions (the dollars) oxidize (they depreciate). In accordance with the ions nature (eurocents or American cents), the resulted product (the exchange) can be released (the exchange takes place) or can be stored on the electrode (the exchange does not take place).

In order to check up the reactions (the currency exchanges) in the electrolysis installation (in the economic environment) different pairs of materials may be chosen (between different currencies) for electrodes (currencies that suffer a currency exchange). In the same way, the type of the salt in the electrolyte may be selected (a certain commission percentage) in order to promote (to facilitate a certain currency exchange: euro vs. dollar) instead of another one (euro vs. rouble). The electrolyte with salts (the commission) contains ions (percentages) that conduct the electrical current (that establish the currencies stock).

In conclusion, the galvanization of the metallic parts (the settlement of the official quota of the currencies units) settled against the corroding process (against monetary depreciation) is equivalent with the protection of an alterable metal (a volatile currency) against the corroding process (against depreciation) due to a deposit through the electrolysis (currency exchange) of an inalterable metal (a stable currency). The object (the currency) to be covered (to be stabilized) is connected on the negative pole of a generator (currency stock) and put in a electrolytic bath (Fig. 4), the electrochemical process (that means the currency exchange) carrying out the role of purification ("stabilizer"), as it is the case of galvanic "covering" of an electrode (or the part to be covered) (Fig. 4,a).

The determination of the official currency parity (O.C.P), taken by twos expresses how many unities from the Y currency may be bought with a unity from the Z currency. The calculation formula is the following (for X = euro, Y = RON and Z = dollar):

$$1 X_{€} = aY_{RON}$$
$$1 X_{€} = bZ_{\$}$$
$$OCP_{RON} = \frac{aY_{RON}}{bZ_{\$}} = cY_{RON} / 1Z_{\$} \qquad (8)$$

where: X, Y, Z are currencies, and *a, b, c* are the proportion of the currencies.

In order to foresee the electrochemical reactions (the currencies quotations) the standard potential (the international currency purchasing power) is used. The standard potential for the currencies shall be determined starting from the official currency parity, calculated by the formula (8). In the Table 1 there are displayed some examples of standard potentials of the chemical elements (equivalent with the international currency purchasing power starting from the currencies official quotations) at 25°C (in the area of monetary stability). Practically, the elements that form the electrodes (the currencies) are classified (quoted) in accordance with their standard potential $E_o$ (international currency purchasing power) like of the normal hydrogen electrode potential in the case of elements (Table 1).

The standard potential (the currency purchasing power in the conditions of international monetary stability) gives the capacity (quotation) in report with Euro that corresponds to the hydrogen from the physically reference electrode potential (Euro is taken as reference) of giving in electrons (in order to become a liquid currency – that has an ample quantity for the currency exchange). The other elements (currencies) have a positive (higher) standard potential (a purchasing power) or a negative one (lower than euro).

In conclusion, the elements with negative standard potential (currencies with a purchasing power under euro) are more exposed to be oxidized (to be depreciated) than the elements with a positive $E_0$ (having a higher purchasing power).

## 5. Conclusions

The currency exchange process with commission may be modeled on the basis of its analogy with the process of ionic exchange between a solution with electrolytes and metallic electrodes which the solution is into contact within an electrochemical system of the type of power chemical source or electrolysis with metal deposit (galvanoplastics).

In the present work it is shown that the electrode potential $E_o$ from the electrode-electrolyte contact represents, from the economic point of view, the sale attractiveness of the currency (assimilated with the metallic electrode) – therefore with its international purchasing power.

Considering the currency euro as the reference electrode, a classification of currencies can be achieved in accordance with the international currency purchasing power, similar to the classification of the electrode potentials $E_o$ of the chemical elements (metals) used as electrodes in the electrochemical systems taken into report with the standard potential of the hydrogen chosen as a reference element (with $E_o = 0$). As in the case of the elements used as electrodes in electrochemistry, the currencies may have a positive potential (a higher purchasing power) or a negative one (with a purchasing power lower) that a quntity of *a* Euro that can to be "oxidyzed" or to suffer "reducing" reactions (that means it suffers depreciation), in comparison with the currencies that have a positive potential $E_o$.